\documentclass[aps,prb,twocolumn,superscriptaddress,showpacs,longbibliography]{revtex4-1}

\UseRawInputEncoding
\usepackage{amsmath}
\usepackage{graphicx}
\usepackage{amsfonts}
\usepackage{verbatim}
\usepackage{dsfont}
\usepackage{color}
\usepackage{hyperref}
\hypersetup{colorlinks = true, urlcolor = blue, linkcolor = blue, citecolor = blue}

\begin{document}

\title{Quantum phase transitions in superconductor--quantum-dot--superconductor Josephson structures with attractive intradot interaction}

\author{Yu-Chun Hsu}
\affiliation{Department of Electrophysics, National Chiao Tung University, Hsinchu, Taiwan}
\author{Wu-Jing Chen}
\affiliation{Department of Electrophysics, National Chiao Tung University, Hsinchu, Taiwan}
\author{Chien-Te Wu}
\affiliation{Department of Electrophysics, National Chiao Tung University, Hsinchu, Taiwan}
\affiliation{Physics Division, National Center for Theoretical Sciences, Hsinchu, Taiwan}

\begin{abstract}

We theoretically study the superconducting proximity effect in a quantum dot coupled to two superconducting leads
when the intradot interaction between electrons is made attractive.
Because of the superconducting proximity effect, the electronic states for the embedded quantum dot are either spin-polarized states with an odd occupation number or BCS-like states with an even occupation number. 
We show that in the presence of an external magnetic field, the system can exhibit quantum phase transitions of fermion parity associated with the occupation number.
In this work, we adopt a self-consistent theoretical method  to extend our considerations beyond the so-called superconducting atomic limit in which the superconducting gap for the leads is assumed to be the largest energy scale. The method enables us to numerically investigate the electronic structure of the dot as results of the attractive interaction. 
For energy phase diagrams in the regime away from the atomic limit, we find a reentrant behavior where a BCS-like phase of the dot exists in 	an intermediate range of the hybridization strength between the quantum dot and the leads.
We also consider Josephson current phase relations and identify a number of examples showing $0-\pi$ phase transitions that may offer important switching effects. 
\end{abstract}
\maketitle
\vskip5mm

\section{Introduction}
\label{intro}

%
%
%
%
%

Introducing localized magnetic impurities into a host superconductor (SC)
leads to the formation of the so-called Yu-Shiba-Rusinov 
(YSR) states~\cite{yu,Shiba,rusinov}. In such a system, quantum phase transitions (QPTs)
associated with fermion parity switches of the YSR states
can be achieved via tuning experimental knobs such as
gate voltages and external magnetic fields~\cite{wu2018,Corral2020}.
Recently, there has also been interest in topological
properties of a chain of magnetic impurities on the surface
of a superconductor
because Majorana bound states, which are regarded as candidates of fault-tolerant quantum computers,
may emerge in such a solid-state system~\cite{glazman,klinovaja}.
Therefore, exploring the physics behind the QPTs of YSR states
can help researchers to advance quantum information science.
A similar and closely related setup 
that also enables one to study the physics of QPTs is
a tunnel junction that consists of quantum dots (QDs) and
two superconducting electrodes.
The latter is the main focus of this work.

Because
physical properties of QDs can be easily tuned 
by varying their sizes, shapes, electron densities, 
as well as electrode voltages, solid-state systems 
containing QDs have been proved
to have useful applications 
in many fields including biomedical applications~\cite{WAGNER201944}
and quantum information technology~\cite{qubit}.
In the latter, qubits, 
the building blocks of a quantum computer, 
are implemented by charge/spin degrees of freedom in QDs. 
Owing to size-tunable emissions of QDs, 
their optical properties are suitable for
fluorescence biomedical applications.
Quantum dots can also be used to build single electron transistors~\cite{SET} 
because of its pronounced Coulomb blockade effect originating
from the presence of a strong Coulomb repulsion.

The physics of heterostructures composed of 
both superconductors and quantum dots has  
become an important and exciting research topic. 
When a quantum dot is in contact with superconducting electrodes, 
the electronic structure of the quantum dot is drastically modified 
due to the local formation of Cooper pairs via the superconducting proximity effect. 
The quantum dot thus can possess BCS-like states 
in contrast to their original discrete states~\cite{beenakker},
which are under a more direct influence of the Coulomb blockade effect.
Furthermore, one of the most important consequences of the SC-QD coupling is the emergence of 
Andreev bound states (ABS),~\cite{meng2009,Bauer2007,YI20131127} 
carrying important information on phase transitions of the dot in SC-QD heterostructures.

Important  physical quantities 
such as the spectral weights~\cite{Bauer2007} and Josephson currents~\cite{Benjamin2007} 
of the SC-QD heterostructures
have been  studied in the literature. In Ref.~\onlinecite{Bauer2007}, 
the superconducting proximity effect on the local spectral properties 
of the dot is investigated. It is found that the low-energy spectrum 
is determined by the superconducting gap of the leads. The situation depicted there 
is similar to a Kondo impurity embedded in a superconductor. 
Consequently, the Kondo effect plays an important role 
in SC-QD heterostructures and it is necessary to compare 
the energy scales of the superconducting gap, $\Delta$, 
and the Kondo temperature, $T_K$. 
As discussed in Refs.~\onlinecite{Bauer2007,delagrange}, 
for cases where $\Delta\ll T_K$, the ground state of the dot 
is a Kondo/BCS singlet state due to the assistance of the Kondo effect. 
In this regime, the Kondo coupling between 
the superconducting electrodes and the quantum dot 
helps the establishment of superconducting correlations in the dot. 
However, as demonstrated in Ref.~\onlinecite{zonda2016}, 
the ground state is a BCS singlet with weak and repulsive Coulomb interaction 
and crosses over to a Kondo singlet when the repulsive interaction is strong. 
For the other extreme limit, $\Delta\gg T_K$, there is essentially no states 
around the Fermi level because of the large gap. Therefore, the Kondo effect is
 suppressed and the ground state of the dot is a Kramers doublet state 
as long as the time-reversal symmetry is preserved.

Another important aspect of the QD-SC coupling is 
its effect on transport properties of SC-QD-SC junctions. 
Specifically, the Josephson effect evaluated in 
Refs.~\onlinecite{Benjamin2007} and~\onlinecite{choi2004} 
can be used to detect signatures of the phase transitions of the quantum dot. 
As illustrated in the above, a BCS-like state of the dot occurs 
when $\Delta\ll T_K$ and, as a result, the transport of a singlet Cooper pair 
from one of the superconducting electrodes to the other 
does not require a sign flip of the singlet state. 
On the other hand, when $\Delta\gg T_K$, the dot is in the doublet state 
and acts as a single magnetic impurity. A Cooper pair is then affected 
by the magnetic impurity and a negative sign is acquired  
when it is transported from one side to the other. 
It is clear that for the latter case, the associated Josephson current
 also gets inverted and the SC-QD-SC is a $\pi$-junction. 
The doublet-singlet phase transitions can thus be experimentally 
confirmed by measuring the current-phase relations of the junctions.

One simple and elegant model to describe a quantum dot coupled to superconducting leads
is the Anderson impurity model. 
Intradot Coulomb interaction $U$
and the coupling 
between the dot and leads
are two important competing energy scales in the model.
The Coulomb interaction between electrons involves four operators, and, as a result,
the Anderson Hamiltonian can not be simply recast into a bi-linear form. Since the Coulomb
interaction has important implications in transport properties, it cannot be neglected
in the problem. 
For example, when the dot is singly occupied, 
it is unlikely to have another electron flowing through the dot
when the Coulomb interaction  is strong and repulsive.
This phenomenon is known as the Coulomb blockade~\cite{anna2008,eichler2009}.
Nevertheless, there exist several ways in the literature to
 estimate
the contribution from the Coulomb interaction including 
the perturbation expansion in the Coulomb interaction~\cite{anna2008,avishai2001,ishizaka1995}, 
mean-field theory~\cite{avishai2001,vecino2003}, noncrossing approximation (NCA)~\cite{clerk2000}, 
numerical renormalization group (NRG)~\cite{choi2004,karrasch2008,oguri2004,yoshioka1999} or quantum Monte
Carlo and functional methods~\cite{rozhkov1999}.

In Ref.~\onlinecite{wentzell2016}, the authors carefully discuss the physics of SC-QD-SC Josephson junctions 
under the influence of a Zeeman interaction in the dot. 
In particular, they use both functional renormalization group (fRG) methods and the self-consistent 
Andreev bound states theory (SCABS) to study the interplay between the Zeeman field, gate voltage, and 
flux dependence of Andreev levels. They found a very good agreement between these techniques even though 
the computational requirements for the fRG approach are rather high.
The SCABS technique is  adopted in the present paper because
it is numerically less demanding while still offers an elegant way to gain insights 
into the physics in SC-QD-SC hybrids.
Very recently, relevant experimental results 
on the Kondo screening-unscreening transition
are reported~\cite{Corral2020}.
There,
they demonstrated how to use a magnetic field to
tune the QPTs of a quantum
dot coupled to superconducting leads in a transistor geometry.
Also, they found that
the magnetic field gives rise to a reentrant transition due to
the competition between the Zeeman shift of the lowest
spin-polarized level and the reduction of the superconducting
gap.

In some special semiconducting quantum dot devices made from $\rm LaAlO_{3}$ or $\rm SrTiO_{3}$~\cite{Cheng2015,cheng2016,Prawiroatmodjo2017},
the electron-electron interaction can be made attractive
by tuning the gate voltage~\cite{cheng2016}. 
It is also demonstrated that  in carbon nanotubes
the excitonic mechanism brings about an 
attractive interaction between two electrons~\cite{Hamo2016}.
In these special QD devices,
the attractive interaction not only has an impact on the full counting statistics~\cite{Kleinherbers2018}
but also causes a charge Kondo effect~\cite{Fang2017,Tabatabaei2018,Fang2018,Choi2020,Giuliano2020,Fang2014} 
instead of spin Kondo effect.
The attractive $U$ charge Kondo effect
is associated with the fluctuations in degenerate ground states with
different even charge occupations. 
It is illustrated in Ref.~\onlinecite{taraphder1991} 
that both the empty and doubly occupied state
have energies lower than those of the singly occupied states, or equivalently,
the spin-polarized states. This leads to ground state fluctuations in the charge channel 
rather than the spin channel. 
Experimentally, PbTe doped with Tl studied in 
Refs.~\onlinecite{costi2012,matsushita2005,matusiak2009,dzero2005}
is the first material that exhibits pieces of evidence of the charge Kondo effect.
Its direct consequences on transport properties in QDs with normal leads have
also been revealed in Refs~\onlinecite{andergassen2011,cheng2013,koch2007,zitko2006,Placke2018}
where theoretical results including susceptibilities and thermoelectric powers are presented.

The above discussion outlined the importance of considering the Zeeman effect as well as attractive interaction
in SC-QD-SC devices. 
In the absence of the Zeeman effect,
the singlet state is always energetically favorable when compared with
the doublet state for an  attractive Coulomb interaction.
Nevertheless, we show in this work that with the inclusion of an applied magnetic field,
the singlet and the doublet states still energetically compete with each other
even when the intradot interaction is attractive.
We investigate the SC-QD-SC structures that are not necessarily in the superconducting atomic limit 
and obtain several important results in the framework of the SCABS theory.
Under an applied magnetic field, the Zeeman energy split of the quantum dot 
is usually much larger than that of the superconductors in SC-QD-SC junctions, 
because the $g$ factor for a semiconducting quantum dot 
(especially for materials with extremely strong spin-orbit coupling)
is usually much higher 
than that for a superconducting material~\cite{gerven2017}.
In the present work, we, therefore, assume the Zeeman splitting in the superconducting electrodes is negligible.

In the paper, we adopt the perturbative SCABS method developed in Ref.~\onlinecite{meng2009}. We aim  to go beyond the superconducting atomic limit and to study the interplay between a Zeeman field and an attractive $U$ interaction. 
Here we study specifically phase diagrams and Josephson current phase relations. 
We shall consider several relevant parameters including hybridization strengths, 
single-particle energies of an electron in the QD measured from the Fermi surface,  
superconducting gaps and phase differences for the superconducting electrodes, and strengths of 
the Coulomb interaction and Zeeman field. In Sec.~\ref{sec2}, we shall present a general description of the SCABS method. 
The results and relevant discussion will be presented in Sec.~\ref{sec3}. We will summarize the paper in Sec.~\ref{sec4}.

\section{Theoretical Model}
\label{sec2}

\subsection{Hamiltonian}
\label{qd_ham}
We start with
the Hamiltonian of the Anderson impurity model for a quantum dot coupled to two
superconducting electrodes. The Hamiltonian is given by
\begin{equation}
H=\sum_{i=L,R}H_{i}+H_{d}+\sum_{i=L,R}H_{T_{i}},
\end{equation}
where 
\begin{subequations}
\begin{align}
\label{shal}H_{i} & =\sum_{{\bf k}\sigma}\varepsilon_{{\bf k}}c_{{\bf k}\sigma i}^{\dagger}c_{{\bf k}\sigma i}-\sum_{{\bf k}}\left(\Delta_{i}c_{{\bf k}\uparrow i}^{\dagger}c_{-{\bf k}\downarrow i}^{\dagger}+{\rm H.c.}\right),\\
\label{dhal}H_{d} & =\left(\varepsilon_{d}+h\right)d_{\uparrow}^{\dagger}d_{\uparrow}+\left(\varepsilon_{d}-h\right)d_{\downarrow}^{\dagger}d_{\downarrow}-Un_{\uparrow}n_{\downarrow},\\
\label{thal}H_{T_{i}} & =\sum_{{\bf k}\sigma }\left(td_{\sigma}^{\dagger}c_{{\bf k}\sigma i}+{\rm H.c.}\right).
\end{align}
\end{subequations}
In this total Hamiltonian, $H_{i}$ is the Hamiltonian for the superconducting electrodes
with $i=L,R$ denoting the left and right electrodes, respectively.
For an electron in the $i$-th lead with a wavevector ${\bf k}$
and 	spin $\sigma$, its kinetic energy is $\varepsilon_{{\bf k}}$.
$c_{{\bf k}\sigma i}$  and $c_{{\bf k}\sigma i}^\dagger$ are the annihilation and
creation operators of the electron, respectively.
$\Delta_{i}$ is the superconducting
order parameter of the $i$-th lead, and $\rm{H.c.}$ 
is the Hermitian conjugate.
$H_{d}$ is the QD Hamiltonian, and we consider only a single energy level, $\varepsilon_{d}$, for the dot. 
$d_{\sigma}$ ($d_{\sigma}^\dagger$) is the annihilation (creation) operator
of a dot electron with spin $\sigma$. $U$ is the Coulomb interaction
between two electrons in the orbital of QD,
$h$ is a
Zeeman interaction, and $n_{\sigma}=d_{\sigma}^{\dagger}d_{\sigma}$ is
the number operator of the dot level with spin $\sigma$. 
Note here that 
$U>0$ denotes an attractive interaction in our convention.
 $H_{T_{i}}$ is
the interaction between the quantum dot and superconducting
leads, and $t$ is the corresponding coupling strength
directly related to the superconducting proximity effect. 

Here we consider an experimentally more relevant situation~\cite{Corral2020}
where the leads are identical $s$-wave BCS superconductors
with a possible nonzero relative phase.
Therefore, both of them have the superconducting order parameter
 of the form $\Delta_{i}=\Delta e^{i\phi_{i}}$,
where $\Delta$ is a constant isotropic gap and $\phi_i$ is its phase. The phase difference
between the leads is denoted by $\phi\equiv\phi_{L}-\phi_{R}$. 
We assume that the density
of states (DOS) in an energy interval $\left[-D,D\right]$ that is of interest is
a constant and specified by $\rho=1/(2D)$. Thus, the total number of states
in the energy interval is independent of the choice of the bandwidth, $D$.
Finally,
the coupling strength $t$ is a real number and is the same for both leads.

As mentioned in Sec.~\ref{intro}, the Coulomb interaction $U$
plays an important role in determining  physical properties of 
SC-QD hybrids. 
When the superconducting gap is much larger than all the
other energy scales in the problem including $U$, 
the hybrids are considered to be in  
the superconducting atomic limit.
Although it is straightforward to 
compute relevant physical quantities  in the atomic limit, 
the superconducting gap in a realistic situation
is usually comparable to
the energy of the atomic level in QD. Therefore, 
it is crucial to study cases where the gap is finite 
while varying other energy scales such as the Zeeman energy and the
Coulomb interaction.
In view of this, 
we follow closely the perturbation method 
presented in Refs.~\onlinecite{meng2009,wentzell2016},
where the Green's function technique is employed and
the superconducting atomic limit 
is treated as the
unperturbed situation. 
In addition, it is shown in Refs.~\onlinecite{meng2009,wentzell2016}
that the results obtained from the SCABS technique agree rather well
with those from the heavy numerical fRG calculations.

In this work, our aim is to investigate the superconducting
proximity effect on  physical behaviors of SC-QD hybrids.
To do so, we first define the quantum dot Green's function 
$G_{dd}(t,t')=-\langle T_{t}\Psi_{d}(t)\Psi_{d}^{\dagger}(t') \rangle$,
where $\Psi_d=\left(d_\uparrow,d_\downarrow^\dagger\right)^T$.
In order to simplify the calculation of finding the Green's function, 
we use
the Matsubara imaginary time formalism 
and the fact that $G_{dd}\left(t,t^\prime\right)=G_{dd}\left(t-t^\prime,0\right)$ to
set $t^\prime\rightarrow 0$ and  $t-t^\prime\rightarrow \tau$. We then have
\begin{align}
G_{dd}(\tau)&=- \langle T_{\tau}\Psi_{d}(\tau)\Psi_{d}^{\dagger}(0) \rangle \\\nonumber
&=-\left[\begin{array}{cc}
\langle T_{\tau}d_{\uparrow}(\tau)d_{\uparrow}^{\dagger}(0) \rangle & \langle T_{\tau}d_{\uparrow}(\tau)d_{\downarrow}(0) \rangle\\
\langle T_{\tau}d_{\downarrow}^\dagger(\tau)d_{\uparrow}^{\dagger}(0) \rangle& \langle T_{\tau}d_{\downarrow}^\dagger(\tau)d_{\downarrow}(0) \rangle
\end{array}\right].
\end{align}	
Next, we use Fourier transformation and the Heisenberg equation of motion to find the Green's function 
in the frequency domain. 
It is written as
\begin{equation}
\label{Gdd}
\hat{G}_{dd}^{-1}(i\omega_{n})=i\omega_{n}+h-\varepsilon_{d}\hat{\sigma}_{z}-t^{2}\sum_{i=L,R}\sum_{{\bf k}}\sigma_{z}\hat{G}_{{\bf k{\bf k}}i}\sigma_{z},
\end{equation}
where $\omega_n$ is the fermionic Matsubara frequency and $\hat{G}_{{\bf k{\bf k}}i}$ is the bare Green's function of the BCS
Hamiltonian of the lead $i$.
It is written as 
\begin{align}
\hat{G}_{{\bf k{\bf k}}i} & =\left(i\omega_{n}-H_i\right)^{-1}=\left(\begin{array}{cc}
i\omega_{n}-\varepsilon_{{\bf k}} & -\Delta e^{i\phi_{i}}\nonumber\\
-\Delta e^{-i\phi_{i}} & i\omega_{n}+\varepsilon_{{\bf k}}
\end{array}\right)^{-1}\\
 & =\left(\begin{array}{cc}
\frac{i\omega_{n}+\varepsilon_{{\bf k}}}{\left(i\omega_{n}\right)^{2}-E_{{\bf k}}^{2}} & \frac{\Delta e^{i\phi_{i}}}{\left(i\omega_{n}\right)^{2}-E_{{\bf k}}^{2}}\\
\frac{\Delta e^{-i\phi_{i}}}{\left(i\omega_{n}\right)^{2}-E_{{\bf k}}^{2}} & \frac{i\omega_{n}-\varepsilon_{{\bf k}}}{\left(i\omega_{n}\right)^{2}-E_{{\bf k}}^{2}}
\end{array}\right),
\end{align}
where $E_{\bf k}=\sqrt{\varepsilon_{\bf k}^2+\Delta^2}$. Note here that we have temporarily
suppressed the Coulomb interaction in deriving Eq.~(\ref{Gdd}).

We use the assumption that the density of states of the leads is a constant $\rho$ to
perform the momentum sum in Eq.~(\ref{Gdd}).
The relation $\hat{G}_{dd}^{-1}=i\omega_n-H_{eff}^0$ allows us to identify the effective 
Hamiltonian of the dot. It is given by
\begin{equation}
\label{heff}
H_{eff}^0=\left(\varepsilon_d+h\right)d_{\uparrow}^\dagger d_{\uparrow}+\left(\varepsilon_d-h\right)d_{\downarrow}^\dagger d_{\downarrow}
-\Gamma_{\phi} \left(d_{\uparrow}^{\dagger}d_{\downarrow}^{\dagger}+\rm{H. c.}\right),
\end{equation}
where $\Gamma_\phi=\Gamma\frac{2}{\pi}\arctan\left({\frac{D}{\Delta}}\right)\cos\left(\frac{\phi}{2}\right)$ and $\Gamma=2\pi t^2\rho$. 
In deriving the effective Hamiltonian, we have set $\Delta\gg\omega_n$,
a consequence from the superconducting atomic limit.
Finally, when the Coulomb interaction are taken into account, we obtain
the full local effective Hamiltonian,
\begin{align}
\label{hameff}
H_{eff}=&\left(\varepsilon_d+h\right)d_{\uparrow}^\dagger d_{\uparrow}+\left(\varepsilon_d-h\right)d_{\downarrow}^\dagger d_{\downarrow}
-\Gamma_{\phi} \left(d_{\uparrow}^{\dagger}d_{\downarrow}^{\dagger}+\rm{H. c.}\right)\\\nonumber
-&\frac{U}{2}\sum_\sigma \left(d_{\sigma}^\dagger d_\sigma-1\right)^2.
\end{align}

\subsection{Spectrum of the effective Hamiltonian}  
\label{spec}

In this subsection, we wish to determine the eigenstates of the effective Hamiltonian, Eq.~(\ref{hameff}), and  corresponding energy eigenvalues.
We first discuss the four eigenstates for a single orbital quantum dot in the absence of superconducting leads.
We use $\left|0\rangle\right.$ to denote a situation where the orbital is not occupied (empty state).
Note here that the empty state has even fermion parity. The other state
also with even fermion parity is the doubly occupied state (paired state), 
$\left|\uparrow\downarrow\rangle\right.$.
The remaining two singly occupied states, which have odd fermion parity, 
are denoted as the spin-up state, $\left|\uparrow\rangle\right.$, and the spin-down 
state, $\left|\downarrow\rangle\right.$.
When superconducting leads were to be absent,
the aforementioned four states would be
the eigenstates of the single orbital quantum dot.
However, when the dot is coupled to superconducting leads, 
it is clear that the empty and paired states themselves alone
are no longer  eigenstates of the effective Hamiltonian, Eq.~(\ref{hameff}).
As shown below, the superpositions of the empty and paired states 
similar to that of the  variational BCS ground state are
the eigenstates. 
By employing the following Bogoliubov transformation,
\begin{subequations}
\label{bcslike}
\begin{align}
\left|+\rangle\right.&=u\left|\uparrow\downarrow\rangle\right.+v\left|0\rangle\right.,\\
\left|-\rangle\right.&=-v\left|\uparrow\downarrow\rangle\right.+u\left|0\rangle\right.,
\end{align}
\end{subequations}
we obtain
\begin{subequations}
\begin{align}
u=&\frac{1}{\sqrt{2}}\sqrt{1+\frac{\xi_{d}}{\sqrt{\xi_{d}^{2}+\Gamma_{\phi}^{2}}}},\\
v=&\frac{1}{\sqrt{2}}\sqrt{1-\frac{\xi_{d}}{\sqrt{\xi_{d}^{2}+\Gamma_{\phi}^{2}}}}.
\end{align}
\end{subequations}
For the singly occupied states,  $\left|\uparrow\rangle\right.$ and $\left|\downarrow\rangle\right.$,
their eigenenergies are not the same when the time-reversal symmetry is broken by the Zeeman interaction $h$,
\begin{equation}
E_{\uparrow\downarrow}^{0}=\xi_{d}\pm h,
\end{equation}
where $+$ and $-$ signs correspond to $E_{\uparrow}^0$ and $E_{\downarrow}^0$ for the $\left|\uparrow\right.\rangle$ and $\left|\downarrow\right.\rangle$ states, respectively.
For the BCS-like states,  $\left|+\rangle\right.$ and $\left|-\rangle\right.$ in Eqs.~(\ref{bcslike}), their eigenenergies are given by
\begin{equation}
E_{\pm}^{0}=-\frac{U}{2}\pm\sqrt{\xi_{d}^{2}+\Gamma_{\phi}^{2}}+\xi_{d},
\end{equation}
respectively.

In order to find ground state phase transitions, we first note that $E_{-}^{0}$ ($E_{\downarrow}^{0}$) is always less than $E_{+}^{0}$ ($E_{\uparrow}^{0}$).
Therefore, the ground state is either the $\left|\downarrow\rangle\right.$ state or the $\left|-\rangle\right.$ state.
Therefore, we only need to compare $E_{-}^{0}$ with $E_{\downarrow}^{0}$ and  phase transitions occur
when $E_{\downarrow}^{0}=E_{-}^{0}$. The equation below characterizes a phase boundary.
\begin{align}
\label{boundary}
\sqrt{\xi_{d}^{2}+\Gamma_{\phi}^{2}}=-\frac{U}{2}+h.
\end{align}
It is obvious that when the Coulomb interaction is attractive ($U>0$),
$h$ cannot be zero in order for the system to transition between
the spin-polarized state, $\left|\downarrow\rangle\right.$, and BCS-like state, $\left|-\rangle\right.$, which is consistent
with the experiment~\cite{Prawiroatmodjo2017}.

\subsection{Perturbation expansion}   
\label{per}
In Sec.~\ref{spec}, we take the limit that the energy gap $\Delta$ is infinite whereas in reality, the energy
gap is usually a few kelvins for most conventional $s$-wave superconducting materials. To incorporate this,
we adopt the formalism developed in Ref.~\onlinecite{meng2009} to consider situations where
other energy scales such as $U$ and $h$ are comparable to the gap. 
Since the details of the SCABS have  already been reported in the literature,
we shall not reproduce them here and only write down the central equations.
As previously discussed, we regard the effective Hamiltonian, Eq.~(\ref{hameff}),
in the atomic limit as the unperturbed Hamiltonian. As a result, the action can
be separated into the local effective action corresponding to Eq.~(\ref{hameff})
and the action of the perturbation that contains terms not included in Eq.~(\ref{hameff}).

The energy corrections $\delta E_s$ (where $s=+,-,\uparrow,\downarrow$) to these four levels 
can thus be obtained from the SCABS theory and they are given by
\begin{subequations}
\label{renorm}
\begin{align}
&\delta E_{\uparrow\downarrow}  =-t^{2}\sum_{\bf k}\left[\frac{1}{E_{\bf k}+E_{+}^{0}-E_{\uparrow\downarrow}^{0}}+\frac{1}{E_{\bf k}+E_{-}^{0}-E_{\uparrow\downarrow}^{0}}+\right.\nonumber\\
&\left.\frac{2\Delta}{E_{\bf k}}uv\left|\cos\frac{\phi}{2}\right|\left(\frac{1}{E_{\bf k}+E_{+}^{0}-E_{\uparrow\downarrow}^{0}}-\frac{1}{E_{\bf k}+E_{-}^{0}-E_{\uparrow\downarrow}^{0}}\right)\right],\\
&\delta E_{+}  =-t^{2}\sum_{\bf k}\left[\frac{1}{E_{\bf k}-(E_{+}^{0}-E_{\uparrow}^{0})}+\frac{1}{E_{\bf k}-(E_{+}^{0}-E_{\downarrow}^{0})}-\right.\nonumber\\
&\left.\frac{2\Delta}{E_{\bf k}}uv\left|\cos\frac{\phi}{2}\right|\left(\frac{1}{E_{\bf k}-(E_{+}^{0}-E_{\uparrow}^{0})}+\frac{1}{E_{\bf k}-(E_{+}^{0}-E_{\downarrow}^{0})}\right)\right]\nonumber\\
&-2\left|\Gamma_{\phi}\right|uv,\\\nonumber
&\delta E_{-}  =-t^{2}\sum_{\bf k}\left[\frac{1}{E_{\bf k}-(E_{-}^{0}-E_{\uparrow}^{0})}+\frac{1}{E_{\bf k}-(E_{-}^{0}-E_{\downarrow}^{0})}+\right.\nonumber\\
&\left.\frac{2\Delta}{E_{\bf k}}uv\left|\cos\frac{\phi}{2}\right|\left(\frac{1}{E_{\bf k}-(E_{-}^{0}-E_{\uparrow}^{0})}+\frac{1}{E_{\bf k}-(E_{-}^{0}-E_{\downarrow}^{0})}\right)\right]\nonumber\\
&+2\left|\Gamma_{\phi}\right|uv,
\end{align}
\end{subequations}
where $E_{\bf k}=\sqrt{\varepsilon_{\bf k}^2+\Delta^2}$. The perturbed energy is then expressed as $E_{s}=E_{s}^0+\delta E_s$.

\section{Results and Discussion}
\label{sec3}

\begin{figure}[h] 
\includegraphics[width=2.7in,clip]{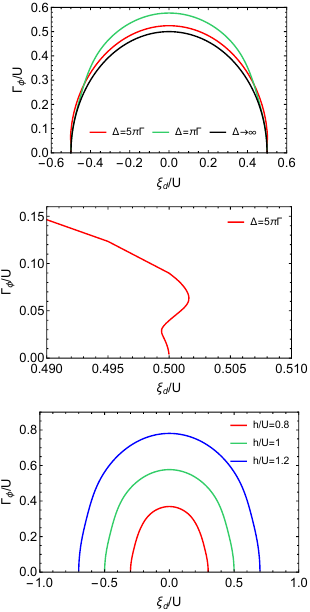}
\caption{
Phase diagrams of a single dot coupled to superconducting leads. The system is in the spin-down (BCS) state below (above) each curve (See main text for details). The bandwidth of the superconducting leads is $D=5\pi\Gamma$. For the top panel, we show the phase diagrams of three different $\Delta$ with a fixed exchange interaction $h=U$. The central panel is the blowup showing the case $\Delta=5\pi\Gamma$ near the edge of the dome in the top panel. In the bottom panel, we consider three different ratios of $h/U$ for a fixed $\Delta/\Gamma=\pi$.
}
\label{fig:1}
\end{figure}

\subsection{Phase diagram} 
In this section, we present our theoretical results on phase diagrams and current-phase relations. We first discuss the phase diagrams. Phase transition lines are  determined by comparing the energies for the four possible states, $\left|+\rangle\right.$, $\left|-\rangle\right.$, $\left|\uparrow\rangle\right.$, and $\left|\downarrow\rangle\right.$. The perturbed energies for these states are  given by $E_{s}=E_{s}^{0}+\delta E_{s}$, where $s=+,-,\uparrow,\downarrow$. As mentioned in Ref.~\onlinecite{meng2009}, the singularities in the integrands of Eqs.~(\ref{renorm}) lead to the limitation of the validity range of the theory.
However, one can extend the range  by using renormalized self-energies instead of bare energies as in the Brillouin-Wigner perturbation theory. To do so, 
we replace $E_s^0$ appeared in the denominators of Eqs.~(\ref{renorm}) by $E_s$. 
As can be seen from the revised expressions, all the energy corrections are now coupled with each other and the corresponding solutions must be determined self-consistently. We find numerically that in the self-consistent scheme $\left|-\rangle\right.$ or $\left|\downarrow\rangle\right.$ is the lowest energy state and thus are competing with each other. 
Accordingly,  phase transition lines are determined from the condition when $E_{\downarrow}=E_{\downarrow}^{0}+\delta E_{\downarrow}=E_{-}^{0}+\delta E_{-}=E_{-}$.

In Fig.~\ref{fig:1}, we present phase diagrams for various situations.
Here, the bandwidth $D$ of the leads is fixed to be $5\pi\Gamma$,
and the mutual interaction between electrons in the dot is made attractive, $-U<0$.
In the top panel, we show the  phase diagrams for three different superconducting gaps
with a fixed Zeeman interaction strength $h=U$. 
The presence of $h$ breaks the time-reversal symmetry, and a Kramer's doublet
is no longer a good eigenstate.
The dot becomes spin polarized when 
the ground state of the system is the spin down state.
This spin-polarized phase corresponds to regions inside the domes in the top panel.
When $\Gamma_\phi$ is large enough, 
the system turns to the BCS-like phase, $\left|-\rangle\right.$, corresponding 
to regions outside the domes due
to a large superconducting proximity effect.
In the atomic limit ($\Delta\rightarrow\infty$),
the radius of the dome
can be directly determined analytically by solving for Eq.~(\ref{boundary}).
(One can also obtain the same result
by considering $D>\Delta\gg\Gamma$ in the SCABS framework.)
Using the self-consistent approach, we
study systems away from the atomic limit
($\Delta$ is finite). It is interesting to note
that the widths of the domes are not changed.
However, the heights of the domes decrease
with increasing $\Delta$. This means
the region for the BCS-like phase shrinks.
This phenomenon  is opposite to the 
results presented in Ref.~\onlinecite{meng2009},
where  for a repulsive intradot
interaction
 the region for the BCS-like phase shrinks when
$\Delta/\Gamma$ is lowered.

In the central panel, we show a blowup
of the $\Delta=5\pi\Gamma$ case of the top panel.
We find an unexpected reentrant behavior
When $\xi_d\gtrsim 0.5U$,
as $\Gamma_\phi/U$ increases from 0, the system
first enters into the spin polarized state and
turns back to the BCS-like state. 
A similar phenomenon is also reported
in ferromagnet-superconductor heterostructures~\cite{wuPRL}
and an atomic Fermi gas~\cite{Chien2006}.
We find that the reentrance is ubiquitous
near $\xi_d\gtrsim 0.5U$ for different 
ratios of $\Delta/\Gamma$
and we chose to show the case of 
$\Delta/\Gamma=5\pi$ because it is more prominent
for illustrative purposes.

In the bottom panel of Fig.~\ref{fig:1}, 
the superconducting gap for the leads is fixed to
be $\Delta=\pi\Gamma$ and the phase boundaries for 
three $h/U$ are shown. For the finite gap, 
the phase transitions
for $\Gamma_\phi/U\rightarrow 0$
occur at the same positions as in the atomic limit of an infinite gap [see Eq.~(\ref{boundary})].
From Eq.~(\ref{boundary}), one can also infer that the applied magnetic field only affects
the radii of domes.
The range for the spin-polarized phase is expanded when the applied magnetic field increases.
It is because the magnetic field tends to break Cooper pairs, and BCS-like states
become unfavorable.

\begin{figure}[h] 
\includegraphics[width=2.7in,clip]{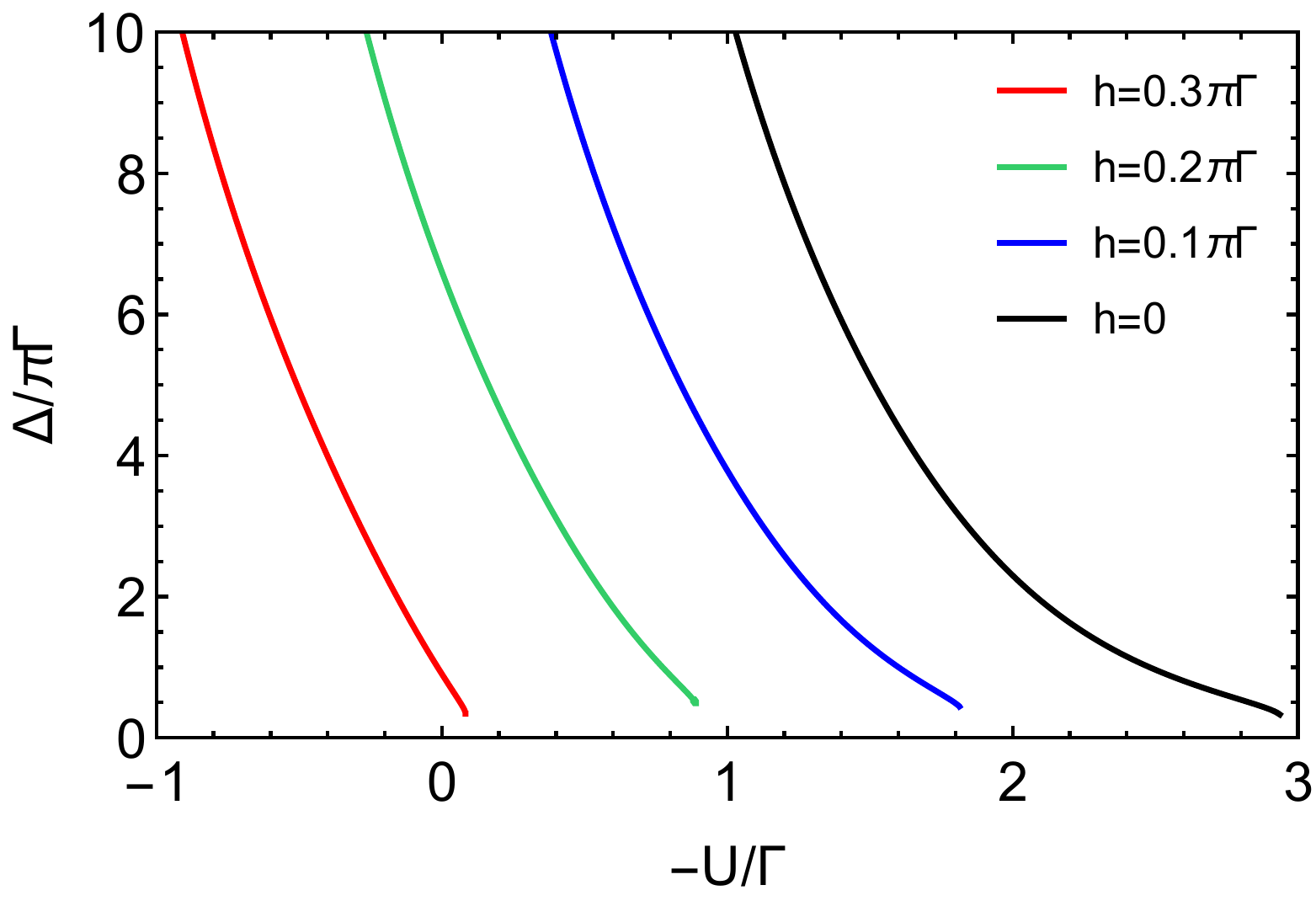}
\caption{
Phase diagrams of a single dot coupled to superconducting leads with a fixed single-particle energy of the bare quantum dot, $\xi_d=0$. 
The bandwidth of the leads is $D=10\pi\Gamma$. Several strengths of an applied magnetic field are considered. The lower left and upper right corners correspond to the BCS-like phase and the spin-polarized phase, respectively (See main text.).
}
\label{fig:2}
\end{figure}

In Fig.~\ref{fig:2}, we fix the single-particle energy of the quantum dot to be at the particle-hole symmetric point, $\xi_d=0$, 
and present phase diagrams by plotting
the superconducting gap versus the Coulomb interaction (we include both attractive and repulsive interaction in the diagrams)
for several $h$.
To the right of each transition line, the Coulomb interaction is either more repulsive or less attractive 
depending on the sign of $U$, and the quantum dot prefers to reside in the single-spin state, $\left|\downarrow\rangle\right.$.
We note that for a given $\Delta$, the phase transition points move to the left
as $h$ increases. 
As in the bottom panel of Fig.~\ref{fig:1}, when the magnetic field is stronger,
Cooper pairs become less stable, and the region for the single-spin state is enhanced.
Therefore, the effect of an applied magnetic field is similar to that of a repulsive Coulomb interaction.
As a result, at a given energy gap, a stronger magnetic field shifts the transition
point to a less repulsive or more attractive Coulomb interaction.
Furthermore, we find that these phase transition curves are
{\it smooth} when the Coulomb interaction
is continuously changed from being attractive to repulsive.
It can also be seen from the
effective unperturbed energies [Eq.~\ref{boundary}] 
that the influences of these two different types of interaction are added together.
In the atomic limit, the Coulomb interaction only shifts the energies of BCS-like states,
and it does not affect single-spin states. Similarly, the magnetic field only shifts
the energies of single-spin states, but not BCS like states in the unperturbed level. Here, we see
that the perturbed energies of the system have the same trend.
Furthermore, we note that
when the applied magnetic field is strong,
the $\left|-\rangle\right.$ state may still be the ground state when the mutual attractive 
interaction is sufficiently large.

\begin{figure}[h] 
\includegraphics[width=2.7in,clip]{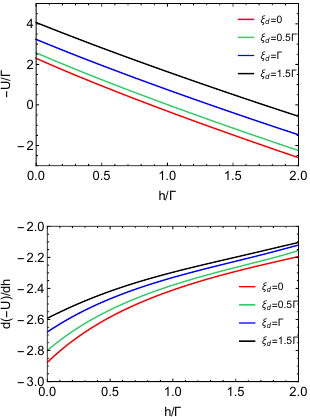}
\caption{
The top panel shows phase diagrams of a single dot coupled to superconducting electrodes with a fixed gap $\Delta=\pi\Gamma$. Four different energy levels, $\xi_d$, of the bare dot are considered. 
For a given $\xi_d$, the area in the upper right (lower left) of each curve corresponds to the spin-down (BCS-like) states. In the bottom panel, we consider the slopes of the curves in the top panel. (See main text.)
}
\label{fig:3}
\end{figure}

\begin{figure}[h] 
\includegraphics[width=2.7in,clip]{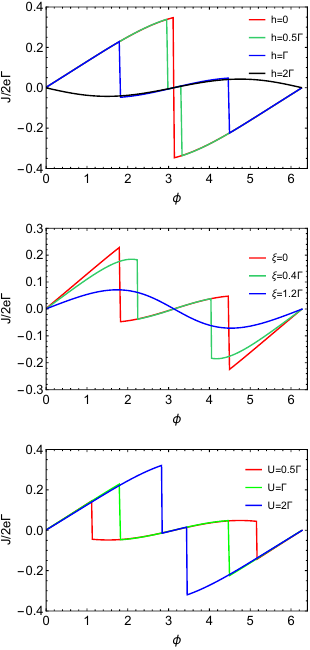}
\caption{
Josephson currents as functions of a relative phase difference, $\phi$, for a single dot coupled to two superconducting leads. The bandwidth of conduction electrons in the superconductors is fixed to be $D=10\pi\Gamma$, and the superconducting gaps are the same for both leads and given by $\Delta=\pi\Gamma$. In the top panel, we consider four different exchange interactions for a fixed mutual intradot interaction $U=\Gamma$ and the bare quantum dot energy level is given by $\xi_d=0$. For the central panel, current phase relations are plotted for three different $\xi_d$ when $h=U=\Gamma$. The bottom panel shows cases for three different Coulomb interactions $U/\Gamma$ for a fixed Zeeman interaction $h=\Gamma$ at the particle-hole symmetric point $\xi_d=0$.
}
\label{fig:4}
\end{figure}

The top panel of Fig.~\ref{fig:3} plots transition lines, between the BCS-like state, $\left|-\rangle\right.$. and spin-down state, $\left|\downarrow\rangle\right.$, for $U$ as functions of $h$ at $\Delta=\pi\Gamma$ for several $\xi_d$. Here, we choose $D=5\pi\Gamma$. To the right of the transition lines, the influence of the Zeeman interaction is strong,  and the ground state is the spin-polarized state for a fixed intradot interaction. For a fixed Zeeman interaction, the system is in the BCS-like regime below the transition lines where the Coulomb interaction $U$ is either repulsive but small or attractive ($-U<0$). It is consistent with the physical picture of a Cooper pair consisting of a pair of electrons bound with each other in that a smaller repulsive $U$ is less detrimental to BCS-like states. For an attractive $U$, the system prefers to be in the BCS-like state. We consider four different $\xi_d$ and find that when it decreases, the BCS-like region shrinks. It is because the system is away from the particle-hole symmetric point and the BCS-like state of the quantum dot becomes more robust  when $\xi_d$ increases. If we consider a given $U$, the system is in the single-spin state to the right of the transition lines where $h$ is large.

From Eq.~(\ref{boundary}), we can see that for a fixed $\Gamma_\phi$ and $\xi_d$,
the transition lines are linear in the atomic limit because $-U/2+h$ is a constant. 
Although  the transition lines in the top panel of Fig.~\ref{fig:3}  appear to be linear 
when $\Delta$ is finite, 
we still compute
their slopes in the bottom panel of Fig.~\ref{fig:3}. 
The results indicate that they deviate from the linear relationship
and show that the system behaves quite differently when
it is away from the atomic limit. In fact, the slope in the atomic limit
is universal regardless of the sizes of $\Gamma_\phi$ and $\xi_d$ and it is always equal to $-2$.
However, we clearly see that only when $h$ is strong, the system behaves as if it is in 
the atomic limit.

From Figs.~\ref{fig:1} to \ref{fig:3}, one can infer that the Coulomb interaction
$U$ and Zeeman effect $h$ both similarly affect phase transition lines. A large and 
repulsive $U$ increases the energy of the BCS-like state, while a high $h$ 
decreases the energy of the single-spin state not only at the unperturbed level
but also the perturbed level to a certain degree. 
As a result, a large and repulsive $U$ requires only a small or vanishing $h$ for the
system to stay in the single-spin (BCS-like) state. The above discussion shows that
it is necessary to go beyond the atomic limit to understand the physics of the SC-QD-SC junctions.

\subsection{Josephson current}   
Next, we discuss Josephson currents in our system. Josephson
currents can be computed by using the formula $J=2e\frac{\partial F}{\partial\phi}$, where
$F=-\frac{\ln(Z)}{\beta}=-\frac{\ln({\rm Tr}(e^{-\beta H}))}{\beta}$ is the free energy and
$\beta=1/ k_BT$. 
For simplicity, we consider the zero temperature limit
 where the free energy is reduced to the ground state energy $E_G$.
Therefore, we first numerically determine the ground state energy 
as
a function of the phase difference $\phi$ between the two superconductors 
in our self-consistent scheme.
The corresponding supercurrent can then be explicitly computed by taking the numerical derivative of $E_G$ with respect to $\phi$.

In the top panel of Fig.~\ref{fig:4}, we show the Josephson current phase relations
for four different Zeeman energies: $h=0$, $0.5\Gamma$, $\Gamma$, and $2\Gamma$. 
When
$\phi\rightarrow0$, $\Gamma_{\phi}=\Gamma\frac{2}{\pi}\tan^{-1}(\frac{D}{\Delta})\cos(\frac{\phi}{2})$
is at its maximum and the energy for the $\left|-\rangle\right.$ state, $E_{-}=E_{-}^{0}+\delta E_{-}$, is in principle at its minimum.
As a result, the ground state energy is usually in the BCS-like phase when $\phi$ is small
or near $2\pi$.
Furthermore, the current phase relation in these $\phi$  regimes is 
given by $J=J_{0}\sin\phi$ corresponding to an ordinary 0-junction.
On the other hand, when $\phi\rightarrow\pi$, $\Gamma_{\phi}=\Gamma\frac{2}{\pi}\tan^{-1}(\frac{D}{\Delta})\cos(\frac{\phi}{2})$
is at its minimum and $E_{-}=E_{-}^{0}+\delta E_{-}$ is higher.
As a result, in a suitable range, the ground state may be the spin-polarized state, $\left|\downarrow\rangle\right.$,
and the current phase relation becomes $J=J_0\sin(\phi-\pi)$
corresponding to a so-called $\pi$-junction.
However, we find that the supercurrent in the $\pi$-junction is small relative to 
that in the 0-junction. 
It is because the spin-polarized state behaves similarly
to a magnetic Kondo impurity that prevents other electrons from passing through
the quantum dot.
In addition, it can be understood by considering 
the superconducting correlations of the dot as discussed in Ref.~\onlinecite{meng2009}.
Because the spin-polarized state always carries a weaker superconducting correlation,
the associated Josephson supercurrent is smaller.

In the top panel of Fig.~\ref{fig:4}, we also find that the $0$-$\pi$
phase transition points are shifted: the region for $\pi$-junction is increased
as $h$ increases.  
We also find that when $\xi_d=0$, the dot is in the 0 phase for the entire range of $\phi$
when $h=0$. 
It is because the $0$ phase corresponds to the BCS-like state and without an 
exchange interaction the dot is never driven to the single-spin state for all possible 
phase difference. 
We note that when $\phi=\pi$, there is a sudden jump from a large positive current
to a large negative current. It suggests that the dot is in the clean limit. 
The sudden jumps may disappear if  we include the contribution from 
the states in the energy continuum
or when the dot is not at the particle-hole symmetric point. 
It is not surprising to see that the $\pi$-junction is not energetically favorable for $h=0$
because
the system cannot stay in the single-spin state  without the inclusion of $h$.
When the Zeeman interaction is strong enough ($h=2\Gamma$),
the spin-polarized state is energetically more favorable for all $\phi$, and the junction turns into
a complete $\pi$-junction.
In Ref.~\onlinecite{rozhkov1999}, it is demonstrated that
when $U$ is strong and repulsive, the Josephson junction
is also a complete $\pi$-junction. 
It confirms the fact that the effect of increasing the Zeeman interaction
is similar to that of increasing the repulsive Coulomb interaction.
This important property can be applied to  nanodevices because switching effects 
here can be easily controlled  by tuning the strength of an external magnetic field. 
On the other hand, the strength of Coulomb interaction is usually material dependent,
and it is more difficult to directly control its associated
switching effects for practical purposes.

In the central panel of Fig.~\ref{fig:4}, we consider additional two slightly larger
bare quantum dot energy, $\xi_d$ at fixed $h$ and $U$. We find that 
as $\xi_d$ increases, the region for the $\pi$ phase is shrinking, and
the BCS-like state is more stable.
The reason behind this is similar to  previous consideration.
When $\xi_{d}$ is away from the particle-hole symmetric point, $\xi_d=0$, 
$\sqrt{\xi_{d}^{2}+\Gamma_{\phi}^{2}}$ becomes larger and $E_{-}=E_{-}^{0}+\delta E_{-}$ becomes lower.
As a result, the BCS-like state (0-phase) is more stable.
For $\xi_d=1.2\Gamma$, 
there is even no $0-\pi$ phase transitions across the entire $\phi$ range.
However, by applying a strong enough magnetic field, the system can still
be driven from the $0$-phase to the $\pi$ phase (not shown) as
clearly demonstrated in the top panel of Fig.~\ref{fig:4}.
The $0$-$\pi$ phase transitions can thus occur 
either by applying a Zeeman field or by tuning the
energy level of the dot via a gate voltage.
Although the system
is a $0$-junction for the entire $\phi$ range when $\xi_d=1.2\Gamma$, 
the size of the supercurrent is smaller compared with the other two cases.
It is because when $\xi_d$ is large, the superconducting correlation
gets weaker, and hence the Josephson current.
In the bottom panel, we consider the particle-hole symmetric point 
and $h=\Gamma$ for several $U$. As can be seen here,
the $\phi$ range for the $\pi$ phase, or the spin-polarized phase, gets
smaller as the attractive interaction gets stronger as we anticipate.

\section{Conclusion}
\label{sec4}

In this paper, we use a relatively simple model to include the local effect 
of an applied magnetic field as well as the phenomenon of 
attractive Coulomb interaction in superconductor-quantum dot-superconductor Josephson junctions.
To go beyond the superconducting atomic limit, we follow a quite successful perturbative scheme
based on the path-integral formalism~\cite{meng2009}. 
In this formalism, all relevant energy scales can be made finite and
thereby suitable for more realistic situations. 

We first present phase diagrams of   superconductor-quantum dot-superconductor  junctions
 under the influence of the interplay between the magnetic field and the attractive Coulomb interaction.
We use a set of self-consistent equations to 
calculate Andreev bound state energies and
Josephson currents as functions of practical experimental knobs
such as the hybridization energy, a phase difference between two superconducting electrodes, 
strengths of Coulomb and exchange interaction, and superconducting energy gaps. 

We show that in the superconducting atomic limit, the effect of applying a magnetic field 
is to shift the energy levels of  single-spin states of the quantum dot by $\pm h$ 
depending on the type of spin. On the other hand, the Coulomb interaction shifts the energy levels 
of BCS-like states (superpositions of vacuum and paired states) by $-U/2$. 
As a result, both the magnetic field and Coulomb interaction 
play crucial roles in determining the phase transition in the atomic limit $\left(\frac{\Delta}{\pi\Gamma}\gg1\right)$.

We find that the system can exhibit reentrant behavior
near phase boundaries in the perturbative scheme
when the system is away from
the atomic limit. 
For an attractive Coulomb interaction, 
the system prefers to reside in a BCS-like state.
For this reason, the system tends to exhibit a
pronounced  superconducting proximity effect in physical quantities
such as superconducting correlation and Josephson supercurrent. 
In order for the dot to transition from the 
BCS-like phase to the spin-polarized phase,
an external magnetic field must be present.
We find that 0-junctions with a higher $\xi_d$, in general, 
have a lower supercurrent. In addition, 
when $U/\Gamma$ ($h/\Gamma$) is
large (small), the BCS-like
regime is enhanced,
and $0$-$\pi$ phase transitions
occur at a higher relative phase between two superconducting leads. 
All the results presented here indicate
that superconductor-quantum dot-superconductor junctions
provide a platform to study quantum phase transitions
as well as switching effects in nanodevices.

\acknowledgements
This work is supported by the MOST Grant No. 108-2112-M-009-004-MY3. 
\bibliography{reference}
\end{document}